# High-precision Distribution of Highly-stable Optical Pulse Trains with Sub-10-fs Timing Jitter

B. Ning, S. Y. Zhang, D. Hou, J. T. Wu, Z. B. Li, & J. Y. Zhao

Department of Electronics, Peking University, Beijing, China

Correspondence and requests for materials should be addressed to: J. Y. Zhao (E-mail: zhaojianye@pku.edu.cn)

High-precision optical pulse trains distribution via fibre links has made huge impacts in many fields. In most published works, the accuracies are still fundamentally limited by some unavoidable noises, such as thermal and shot noise from conventional photodiodes, thermal noise from mixers. Here, we demonstrate a new high-precision timing distribution system by using highly-precision phase detector to overcome the limitations. Instead of using photodiodes and microwave mixers, we use several fibre Sagnac-loop-based optical-microwave phase detectors to realize optical-electrical conversion and phase measurements, for suppressing the noises and achieving ultra-high accuracy. A 10-km fibre link distribution experiment shows our system provides a residual instability at the level of $4.6 \times 10^{-15}$/1-s and $6.1 \times 10^{-18}$/10000-s, with an integrated timing jitter as low as 3.8 fs in a bandwidth of 1 Hz to 100 KHz. This low instability and timing jitter makes it possible that our system can be used in the optical clock distribution or the applications for the facilities which require extremely accuracy frequency/time synchronization.

Distributions of high-precision timing signal (microwave or optical pulse) are necessarily required in many areas of human activity, ranging from practical applications to scientific research. The techniques also directly become the basis in lots of circumstances, *e. g.*, in telecommunication, navigation, astronomy, and physical experiments[1-8]. To share the most precise reference sources which are usually operated at specialized laboratories only, high-precision timing signal distribution techniques are very commonly needed. Especially in some modern large-scale scientific facilities, such as X-ray free-electron lasers (XFELs) and particle accelerators, extremely high timing accuracy of the distribution is required to synchronize the sources in remotely located sites[2, 9-11]. In the past few decades, varieties of RF signal distribution techniques have been developed. One commonly employed way is using satellite methods such as global positioning system and two-way satellite time and frequency transfer[12, 13].

The satellite methods, however, have very moderate performance in accuracy and are unsuitable for most of the modern applications aforementioned. For such applications, current most appropriate and elegant way is using optical fibre links to distribute the RF signals[3]. Research has demonstrated that the fibre-based techniques can provide accuracies and stabilities orders of magnitude higher than the satellite methods[3], as fibre has both low attenuations and high reliabilities. For the fibre-based techniques, lasers play very crucial parts, because they are used as the transmitting sources and directly determine the distribution performances. State-of-art fibre-based RF signal distribution technique can provide stabilities of $10^{-19}$ level over one day[14]. While RF transfer by modulated continuous-wave lasers is a very direct way[14-17], using mode-locked lasers

(MLL) as the sources has attracted more interests recently[2, 18-22]; it can simultaneously provide optical and microwave timing information. With careful optimization[23-24], the MLL can easily generate regularly spaced optical pulse trains with extremely low timing jitter within attosecond scale, which allows it to distribute timing signals with unprecedented precision[2]. Many studies have already been carried on in this area. However, most of the works contain photodiodes and microwave mixers for optical-electrical conversions and phase measurements in some necessary procedures[14-22], which are microwave-to-optic modulations, fibre link stabilisations and RF signal extractions. The devices could induce unavoidable timing noises such as shot noise, thermal noise and amplitude-modulation-to phase-modulation (AM-to-PM) noise[25-27]. The noises obviously reduce the precisions of the laser and limit the distribution accuracy in tens of femtoseconds scales (typical >15 fs for timing jitters and > 30 fs for long-term drift).

In this paper, we demonstrate a new fibre-based optical pulse signal distribution system. Instead of using the traditional electronic devices, in our system, we apply some easily implemented fibre Sagnac-loop-based optical-microwave phase detectors to perform all the aforementioned optic and microwave related procedures, with timing accuracies of several-hundred-as. We tested our system on a 10-km outdoor fibre link, and the results showed that the precision of distribution broke through the limitation of conventional RF transfer[14-22], achieved 3.8 fs integrated timing jitter in a bandwidth of 1 Hz to 100 KHz and 27 fs root-mean-square (rms) timing drift over 12 hours. The residual instability reached $6.1 \times 10^{-18}$ level at 10000-s, which is quite sufficient for long-term transfer of the most advanced clock standards such as optical clocks. Therefore, our frequency distribution technique may provide a very powerful tool for transferring the timing signal of optical clocks without stability loss. This technique may also be used in the application of facilities which necessarily require short-term synchronization of sub-10-fs timing jitter, such as the next-generation of X-ray free-electron lasers.

## Results

**High-precision stabilisation of the femtosecond MLL.** Figure 1a outlines the principles of the high-precision stabilisation of the MLL by locking the laser to a microwave reference via OM-PD. For the stabilisation of the MLL, it is realized by phase-locking the laser to a stable microwave reference with ultra-high accuracy. Conventionally, the phase-locking is achieved by first detecting the phase of the MLL using a photodiode, then phase-comparing the MLL and the reference microwave using a frequency mixer, and at last using the phase-error signal to adjust the length of an intra-cavity piezoelectric transducer (PZT) and therefore control the MLL's phase. This technique, however, suffers from shot and thermal noise generated by the photodiode and the thermal noise of the frequency mixer; it is also limited by the resolution and response time of the PZT. As a result, the accuracy of the stabilisation is limited in tens femtosecond scales. To overcome the limitation, we remove the photodiode and frequency mixer, and use an OM-PD for the phase-comparing. Moreover, to control the MLL's phase much more accurately, we involve a novel technique. Both pump modulation and PZT controlling is applied to ensure long-term, high-accuracy controlling of the laser phase[28]. More information about the technique can be found in Ref. 28 (also see Methods Section).

With the new stabilisation technique for MLL, we successfully phase-locked the MLL to a 6-GHz reference microwave. Residual phase noise[29, 30] of the phase-locking is measured to evaluate the performance. Figure 1b gives the measured data. The phase noise reaches -121 dBc/Hz and -145 dBc/Hz at 1 Hz and 100 kHz offset frequency, respectively. By integrating the phase noise in a bandwidth of 1 Hz to 100 KHz, we have an integrated timing jitter of 1.2 fs. To the authors' knowledge, this is the first time that sub-2-fs timing jitter is achieved in stabilisation of MLLs. The ultra-low timing jitter proves the optical pulses have been stabilised to the microwave reference excellently.

**Experimental setup of ultrastable optical pulse trains distribution system with fibre stabilisation**

Experimental setup of our distribution system with the fibre stabilisation is demonstrated in Figure 2. A classical round-trip fibre stabilisation method is used here to compensate the phase fluctuation induced by acoustic and thermal noise. A 6-GHz microwave generated from a commercial RF signal generator (Agilent, E8257D) is used as the reference microwave. By using a RF signal amplifier and a 1-to-3 power splitter, we obtain three identical microwave reference signals with powers of ~25 mW at the local site. One of the microwaves is used for the stabilisation of the MLL via the OM-PD. One is used to compensate the phase fluctuation of fibre link. The rest is used for the out-of-loop measurement. The laser beam generated from the MLL is split into two beams. One beam with power of ~ 10 mW is phase-detected with a reference microwave in an OM-PD, to generate an error signal, for phase-locking the MLL. The other beam with power of ~30 mW is fed into a fibre link. The fibre link consists of a motorized optical delay line (ODL), a fast PZT-based optical delay line, a 1.1-km dispersion-compensated fibre (DCF), and an 8.9-km commercial single-mode fibre (SMF). The fibre link is installed in spools and directly exposed to an outdoor environment. The total attenuation of the fibre link is ~7 dB. The optical pulses received at the remote site have durations of less than

100 ps. By using a 90:10 fibre coupler and a back reflector, 90% of the received optical beam is reflected back, while 10% of the beam is used for RF signal extraction and measurements. The part with lower power first passes through an (Er-doped fibre amplifier) EDFA to compensate the optical loss from the 90:10 fibre coupler. The optical pulses then pass through a 50:50 fibre coupler. One part with power of ~27 mW is used for out-of-loop performance test by comparing the phase difference between the optical pulse and a part of microwave reference in an out-of-loop OM-PD, while the other part is for the user. It can be used to realize the high-accuracy RF signal extraction or optical-optical synchronization[31-35].

For the 90% reflected optical pulse trains, it is received by using an optical circulator at the local site. A second EDFA is also used here to obtain a high power pulses. The backward pulses are amplified to ~25 mW and fed into an OM-PD to detect the phase fluctuation for the one-round trip. The phase-error signal is filtered by proportional-integral-derivative (PID) module to remove long-term drift and fast noise. Then the PID-regulated error-signal is used to drive the PZT-based optical delay line in a fast speed (~1000 times/s), while the motorized ODL is adjusted at a speed of ~1 times/s to make sure that the fibre length drift is within the locking bandwidth of the PZT-based optical delay.

**Frequency instability and phase noise of optical pulse trains distribution.** Figure 3 shows the optical pulse trains distribution phase noise performance measured at the out-of loop. Similar with most research, the residual phase noise of the distributed pulse timing signal is measured to demonstrate a short-term distribution performance. Figure 3 shows the phase noise with link compensation reaches -104 dBc/Hz and -140 dBc/Hz at 1

Hz and 100 kHz offset frequency, respectively. By integrating the phase noise in a bandwidth of 1 Hz to 100 KHz, we have a short-term integrated timing jitter of 3.8 fs. The timing jitter well reaches sub-10-fs scale. We also test the long-term performance of our system by recording the residual timing drift. Figure 4a shows the timing drift over a 12-hour period of operation. The drift is measured and calculated by using a separated OM-PD and a high-accuracy voltmeter (3 Hz measurement bandwidth, 2 samples/s). A 27 fs rms timing drift is observed, while the drift range could achieve as high as 35 ps (peak to peak) when the compensation system is off. Furthermore, the Allan Deviation calculated from the recorded timing drift data is used to demonstrate the residual instability of the distributed pulse timing signal (filled triangle in Fig. 4b). Figure 4b shows the instability reaches $4.6 \times 10^{-15}$ at 1-s and $6.1 \times 10^{-18}$ at 10000-s averaging time, respectively (3-Hz measurement bandwidth). This residual instability is even lower than H-master clock[36] and optical clock[37].

## Discussion

In this paper, we demonstrate a high-precision fiber-based optical pulse trains distribution system. In the system, we remove the traditional devices such as photodiodes and microwave mixers, while photodiodes could induce shot noise, thermal noise and AM-to-PM noise. Instead, we applied several OM-PDs to perform the procedures of MLL stabilisation, fibre link stabilisations. The OM-PD is built based on the principle of Sagnac loop, and is very robust and reliable. Unlike the research in Ref. 33, our study not only involves fibre stabilisation, but also includes ultralow-noise MLL stabilisation. The timing-stabilised pulse trains at the end-station have many applications, for example, it can be used to synchronize with VCO for RF extraction, mode-lock laser for ultralow noise optical-optical synchronization. Many alternative electrooptic techniques have been demonstrated for RF extraction[31, 32, 34, 35] and optical-optical synchronization[2, 33]. All of these techniques have achieved sub-fs level, which is good enough for signal synchronization.

The results on a 10-km fibre link experiment prove that the accuracy of the ultralow noise optical pulse signal distribution system achieved sub-10-fs scales for short-term (1s), and provided a 30 fs scales for long-term performances (12 hours). To our best knowledge, a long-distance (longer than 10 km), optical pulse signal distribution system with sub-10-fs timing jitter is demonstrated via low-cost SMF link for the first time. Compared with other high-accuracy fibre stabilisation technique, *e. g.*, using periodically poled $KTiOPO_4$ (PPKTP) single-crystal cross-correlators[2, 33], the OM-PD method is much more easily implemented and robust, with quite larger timing delay detection range and lower cost.

The system is potentially suitable to be used in timing synchronization for the next generation of XFELs and particle accelerators, of which different parts need to be distantly located to achieve sufficient accelerating distance while working in tightly synchronized conditions. The facilities are being built all over the world (e. g., Pohang XFEL in Korea, RIKEN/SPring XFEL in Japan and DESY XFEL in Europe), while the schemes which could satisfied their extremely high timing synchronization accuracy are still being developed. The accuracy of the synchronization directly affects the performance of the XFELs in Pohang and DESY test. Moreover, microwaves of different frequencies are required in different locations. The system proposed in this paper does satisfy these requirements. The provided accuracy of the distribution is in tens femtosecond scale, which is sufficient for most XFELs, while the 10-km distribution distance also long enough for most facilities. At the same time, the residual instability of our distribution system is superior to that of the optical clocks. This makes it possible to synchronize the timing signal of optical clock to the remote facilities, when the optical pulses are stabilised to optical clocks. Furthermore, the different harmonics of the pulse trains' repetition rate can be extracted into RF signal with different frequencies simultaneously. This characteristic is valuable for microwaves extraction of different frequencies. The techniques proposed in this work would be applied to the time and frequency distribution over free space in our future work.

## Methods

**The Er-doped fibre MLL.** We use a passively nonlinear polarization rotation (NPR) mode-locked Er-doped fibre as the optical source. The MLL works in the stretched-pulse regime and has a fundamental repetition rate of 173 MHz. Its $35^{th}$ harmonic of repetition rate, 6-GHz, is used for the RF signal extraction and phase measurement to provide a high sensitivity. A 40-cm highly-doped Er gain fibre with cumulative anomalous group velocity dispersion is use, while the other part of the fibre cavity is built using common SMF. For reducing the intra-cavity negative dispersion, we use a space isolator instead of fibre isolator. By optimizing of the ratio of positive dispersion fibre to negative dispersion fibre, we set the intra-cavity dispersion at the close-to-zero dispersion condition to minimize the timing jitter. The MLL is pumped through a 980 nm/1550 nm wavelength division multiplexing (WDM) fibre coupler by a 650-mW, 980-nm diode. A polarization beam splitter, three half-wave and quarter wave plates are to induce the mode-locking. The output power of the MLL is ~50 mW.

**Description of optical-microwave phase detectors (OM-PD).** The detailed description on OM-PD can be found in Ref.34. It was originally proposed in 2004 (Ref.38) and demonstrated in 2012 (Ref. 34) by Jungwon Kim, and the authors tried to apply it for link stabilisation purpose. The OM-PD fundamentally performs the phase detection based on the Sagnac-loop interferometer theory[39]. It consists of a circulator, a unidirectional high-speed $LiNbO_3$ phase modulator and a specially designed polarization-maintaining (PM) fibre Sagnac loop with a short length. When a microwave signal with a frequency of integer-multiple to the laser is applied to the unidirectional phase modulator, the copropagating pulse experiences the phase modulation while the counterpropagating pulse does not. The phase of copropagating pulses is modulated according to the temporal position between the optical pulses and the driving microwave signals. The power difference between the two outputs of the Sagnac loop is proportional to the phase error between the optical pulse trains and the driving microwave signal. A balanced photodetector is used for precise optical-microwave phase detection. The most important feature of OM-PD is the implementation of phase detection is finished in term of optical before the photodetection is involved.

**Stabilisation of MLL via both pump modulation and PZT controlling.** The stabilisation of our MLL is based on both pump modulation and PZT controlling. The principle of using pump modulation to adjust the MLL's phase is that the intra-cavity optic intensity change will lead to linearly changes of the refraction indexes of both Er-doped fibre and SMF. The underlay theory of this principle is that the intra-cavity optic would interact with Er atoms and the silicon in the cavity based on nonlinear effects, and the effects determine that when the optic intensity changes linearly in a very short range, the refraction indexes of the two materials will also change linearly[28]. Considering that the repetition rate $f_r$ of the MLL is determined by the MLL's cavity length $L$ and the cavity's average refraction index $n$ as:

$$f_r = \frac{c}{nL}, \tag{3}$$

where $c$ is the light velocity. It can be seen that changing $n$ will also change $f_r$, therefore change the phase of the laser. Though the adjusting range is short, the advantage of the pump modulation method is that the modulation speed and accuracy is greatly improved when comparing with using PZT only[28].

Therefore, to achieve long-term and high-accuracy, we combined the pump modulation and PZT controlling. While pump modulation is used to achieve fast, high-accuracy stabilisation, long-term stabilisation is ensured by using PZT length adjusting. A commercial current supply (Thorlab, ITC110) is used for the pump modulation, and the PZT (PI, P-840.20) is driven by a 100-V voltage supply and has an adjusting range of 30 um.

**Fibre link stabilisation using both fast and slow feedbacks.** We use both motorized and PZT-based optical delay lines to stabilise the fibre link. The advantage of this method is that the accuracy is very high, while the compensation range is also very large. The PZT-based optical delay line (XMT, 40VS12) is driven by a 150-V voltage supply, with a resolution of 2 fs/V and a compensation range of 200 fs. The motorized optical delay line (General photonics, MDL-002) has an adjusting accuracy of 1 fs and a large compensation range of 500 ps. It is driven slow to make sure that the fast drift is within the compensation range of the PZT-based optical delay line. Both the stabilisation link and the long fibre link work in a reciprocal situation, so by simply locking the phase-error between the local microwave reference and the backward optical pulse trains to near zero, the whole link is stabilised, and optical pulses with stable phases can be received in the remote site. Furthermore, phase fluctuations induced by optical intensity noises could be obviously reduced by detecting the near-zero phase-differences, as in this way the detections are operated when the in-Sagnac-loop signals are almost orthogonal.

## Acknowledgment

This work was supported in part by the Nature Science Foundation of China under Grant Grant 11027404. The authors would like to thank Prof. Zhigang Zhang from the department of electronics, Peking University, for helpful discussion and assistant on the experimental setup of mode-locked laser. The authors appreciate the previous excellent work of Jungwon Kim and the fruitful communication with his group.

## Author contributions

J. Y. Z. and B. N. developed the concept. B. N., D. H. and J. T. W. designed the mode-locked laser and fibre link. B. N., S. Y. Z. and Z. B. L. designed the OM-PDs, B. N., D. H. and J. Z. designed RF signal processing circuits and other electronic servo systems. B. N. and Y. Z. collected the data. Finally, B. N. and D. H. wrote this manuscript. All authors joined the discussion and provide the comments. B. N. and Y. Z. contributed equally to this work.## Additional information

**Competing financial interests:** The authors declare no competing financial interests.

# Figure Legends

**Figure 1 | High-precision stabilisation of the MLL and measured residual phase noise.** (a) Schematic of the high-precision stabilisation of the MLL by locking the laser to a microwave reference via OM-PD. HV: high-voltage; PID: proportional-integral-derivative; LF: low-frequency; PZT: piezo-electric transducer; CM: collimator; WP: wave plate; EDF: erbium-doped fibre; PBS: polarization beam splitter; WDM: wavelength division multiplexer. (b) Measured residual phase noises of the phase-locking. The locking bandwidth is set to be 100 KHz.

**Figure 2 | Experimental setup of optical pulse trains distribution and measurement system.** A stabilised optical pulse trains are distributed to the "remote site" via 10-km fibre link. At the remote site, part of the pulse trains is used for user. The rest is reflected to the "local site" via the same fibre link. An optical-microwave phase detection technique is used to generate a high-precision phase-error signal, for compensating the phase fluctuations in the fibre link.

**Figure 3 | Out-of-loop residual phase noise at the remote site.** (i) out-of-loop residual phase noise of the optical pulse trains distribution system, without link compensation. (ii) out-of-loop residual phase noise of the optical pulse trains distribution system with compensation. (integrated rms timing jitter 3.8 fs [1 Hz–100 kHz]) (iii) noise floor of the distribution system when the fibre link is shorted.

**Figure 4 | Residual instability of the optical pulse trains distribution system.** (a) Timing drift of the optical pulse trains distribution system with compensation (rms timing drift 27 fs, over 12 hours). (b) The measured fractional frequency instabilities. The blue line (filled circle) is the result of the free running fibre link, the red line (filled triangle) shows the result with phase fluctuation compensation, and the pink line (filled diamond) and black line (filled square) are the frequency stabilities of the H-maser and optical clocks used for comparison, respectively.

# Figure

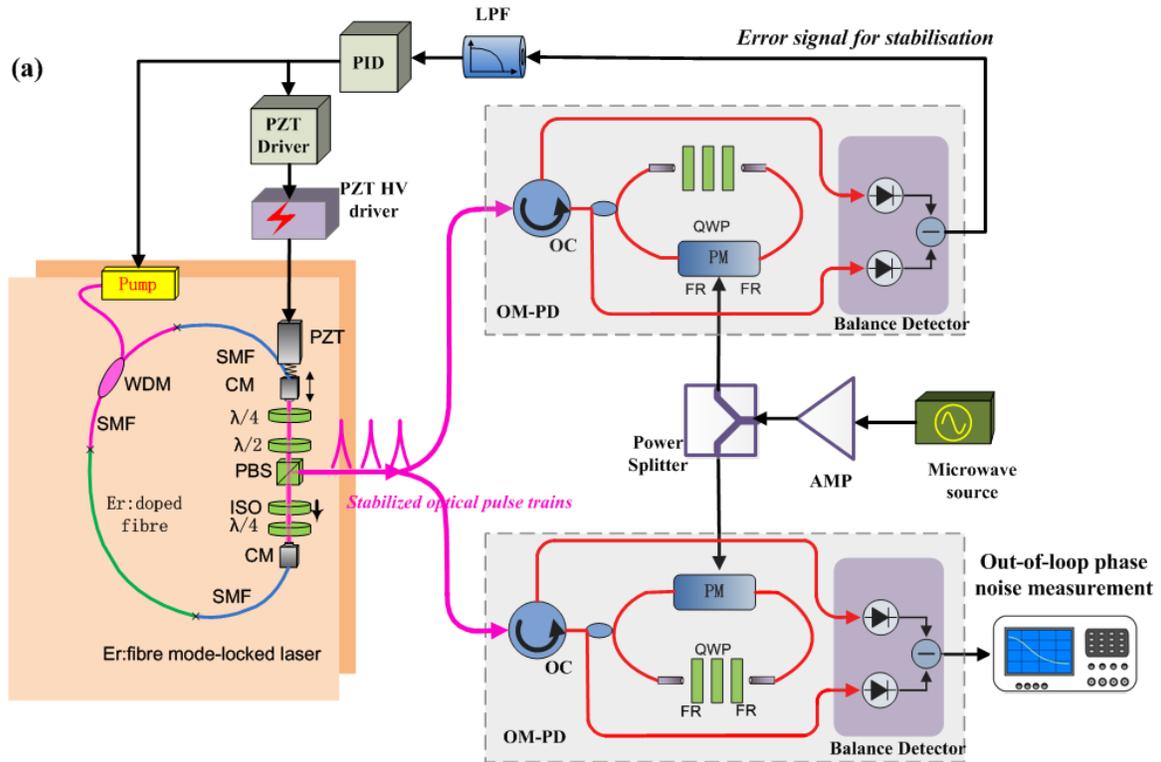

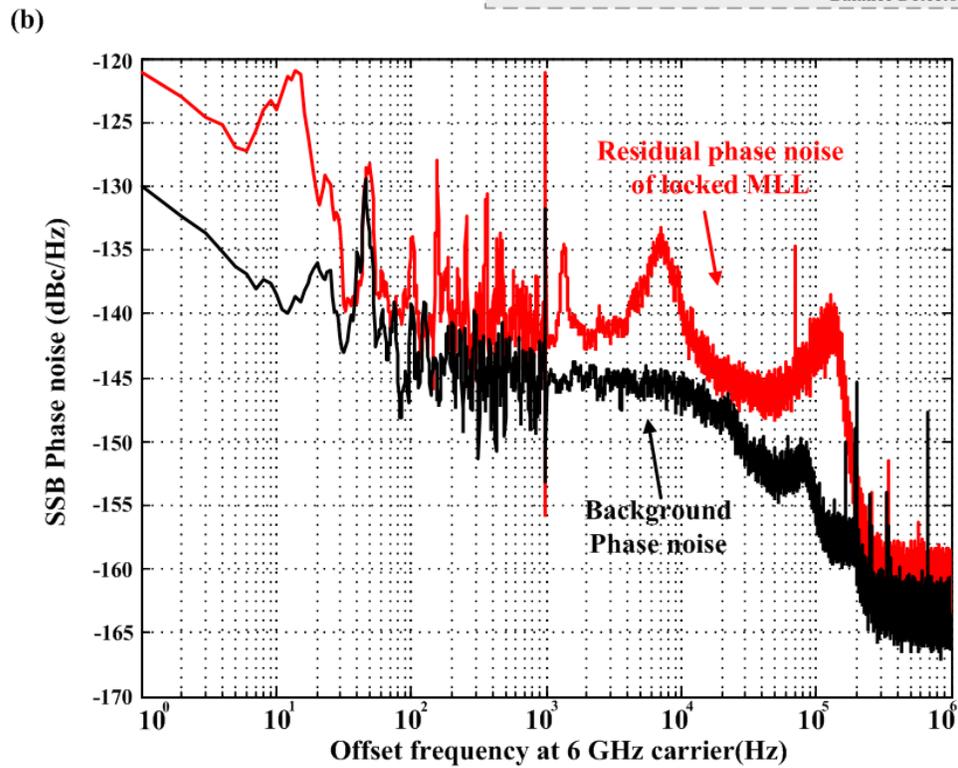

Figure 1

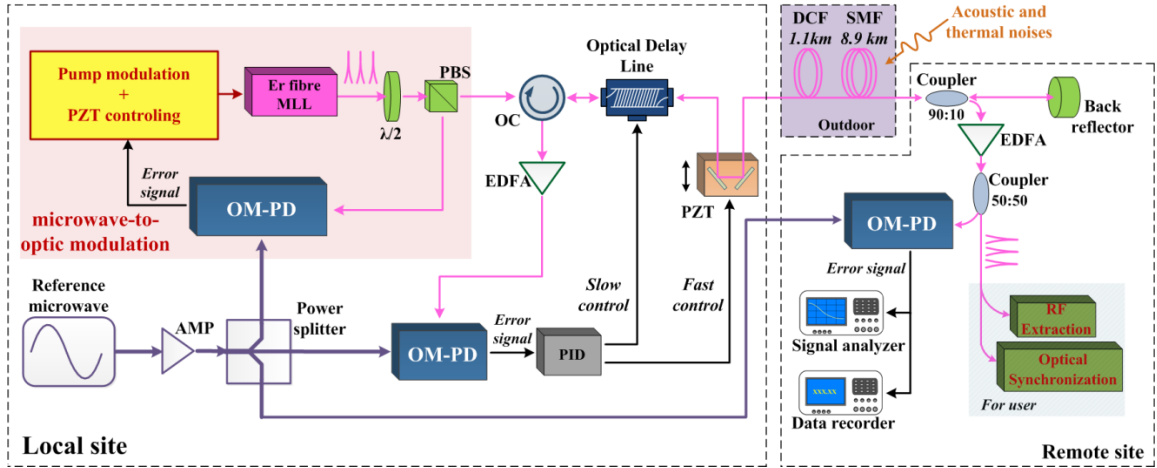

Figure 2

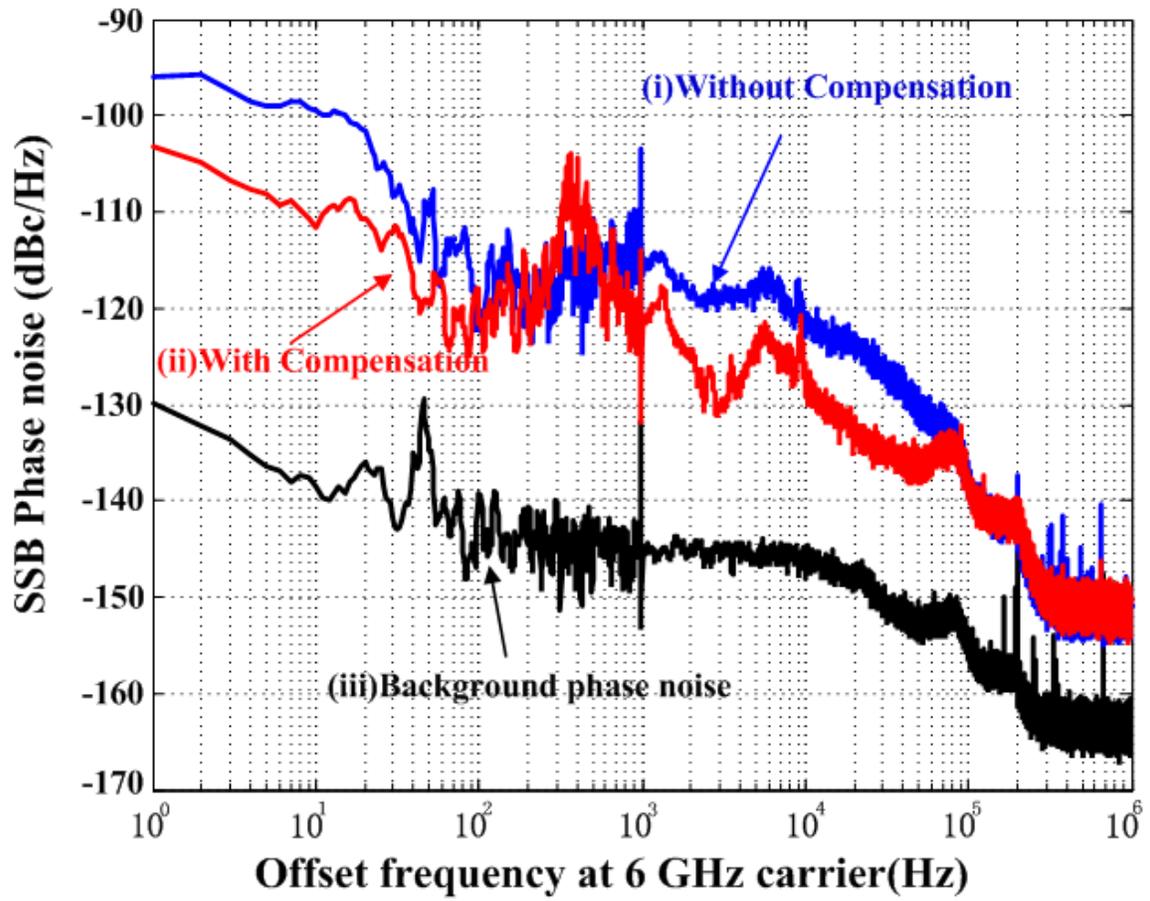

Figure 3

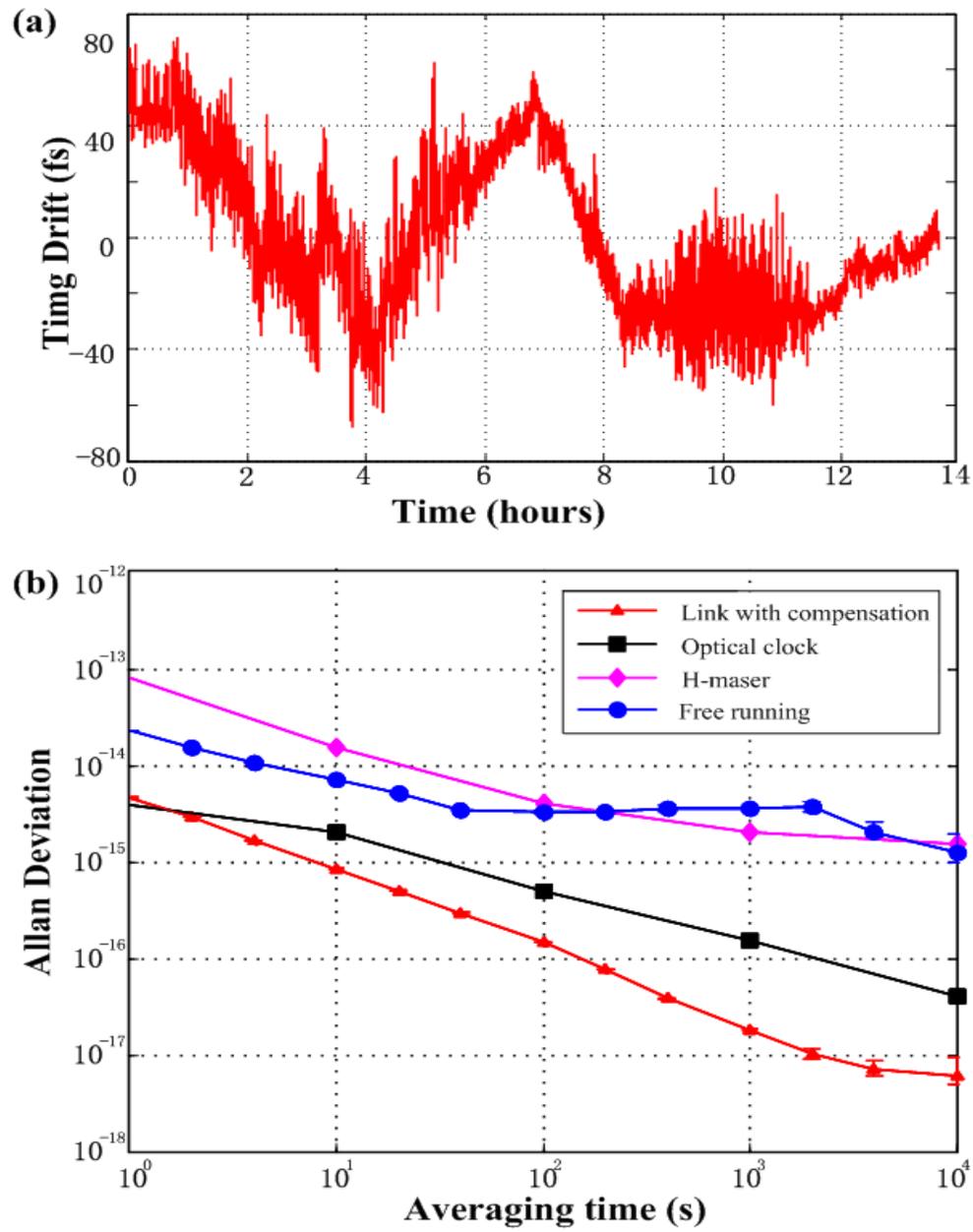

Figure 4